\documentclass[12pt,preprint]{aastex}
\usepackage{amssymb,latexsym,graphics,eufrak}
\begin{document}
\title{Pair winds in Schwarzschild space-time with application to hot bare
strange stars}
\author{A.G. Aksenov~\altaffilmark{1}, M.~Milgrom, V.V.~Usov}
\affil{Center for Astrophysics, Weizmann Institute, Rehovot 76100,
Israel} \altaffiltext{1}{Institute of Theoretical and Experimental
Physics, B.~Cheremushkinskaya, 25, Moscow 117259, Russia;
Alexei.Aksenov@itep.ru}
\begin{abstract}
We consider a time dependent, spherically outflowing wind, in
Schwarzschild space-time, consisting of electron-positron pairs
and photons . Without assuming thermal equilibrium, we account for
the microphysics, including two-body processes ($ee\rightarrow
ee$, $\gamma e\rightarrow \gamma e$, $e^+e^-\leftrightarrow
\gamma\gamma$) and their radiative three-body variants
($ee\leftrightarrow ee\gamma$, $\gamma e\leftrightarrow \gamma
e\gamma$, $e^+e^-\leftrightarrow \gamma\gamma\gamma$).  We present
a finite-difference scheme for solving the general relativistic
kinetic Boltzmann equations for pairs and photons. We apply this
to the concrete example of a wind from a hot, bare, strange star,
predicted to be a powerful source of hard X-ray photons and
$e^\pm$ pairs created by the Coulomb barrier at the quark surface.
We study the kinetics of the wind particles and the emerging
emission in photons and pairs for stationary winds with total
luminosities in the range $10^{34}-10^{39}$ ergs~s$^{-1}$, for
different values of the injected photon-to-pair ratio. The wind
parameters--such as the mean optical depth for photons, the rates
of particle number and energy outflows, bulk velocity, and number
density of the pair plasma--are presented as functions of the
distance from the stellar surface, as well as characteristics of
the emergent radiation. We find that photons dominate in the
emerging emission, and the emerging photon spectrum is rather hard
and differs substantially from the thermal spectrum expected from
a neutron star with the same luminosity. This might help
distinguish the putative bare strange stars from neutron stars.

\end{abstract}
\section{Introduction}
Compact astronomical objects identified with neutron stars,
strange stars, and black holes are believed to be sources of
electron-positron ($e^\pm$) pairs that form in their vicinity by
different mechanisms and flow away. Among these objects are radio
pulsars (Sturrock 1971, Ruderman \& Sutherland 1975, Arons 1981;
Usov \& Melrose 1996; Baring \& Harding 2001), accretion-disc
coronas of the Galactic X-ray binaries (White \& Lightman 1989;
Sunyaev et al. 1992), soft $\gamma$-ray repeaters (Thompson \&
Duncan 1995; Usov 2001b), cosmological $\gamma$-ray bursters
(Eichler et al. 1989; Paczy\'nski 1990; Usov 1992), etc.. The
estimated luminosity in $e^\pm$ pairs varies greatly depending on
the object and the specific conditions: from $\sim
10^{31}-10^{35}$ ergs~s$^{-1}$ or less for radio pulsars up to
$\sim 10^{50}-10^{52}$ ergs~s$^{-1}$ in cosmological $\gamma$-ray
bursters.
\par
For a wind out-flowing spherically from a surface of radius $R$
there is a maximum (isotropic, unbeamed) pair luminosity beyond
which the pairs annihilate significantly before they escape (see
Beloborodov 1999 and references therein). This is given by
\begin{equation}
L_\pm^{\rm max} = {4\pi m_ec^3R\Gamma^2 \over\sigma_{\rm T}}\simeq
10^{36}(R/10^6{\rm cm})\Gamma^2\,{\rm ergs~s}^{-1}, \label{Lpmmax}
\end{equation}
where $\Gamma$ is the pair bulk Lorentz factor, and $\sigma_{\rm
T}$ the Thomson cross section. When the injected pair luminosity,
$\tilde L_\pm$, greatly exceeds this value the emerging pair
luminosity , $L_\pm$ , cannot significantly exceed $L_\pm^{\rm
max}$; in this case photons strongly dominate in the emerging
emission: $L_\pm\lesssim L_\pm^{\rm max}\ll \tilde L_\pm\simeq
L_\gamma$. Injected pair luminosities typical of cosmological
$\gamma$-ray bursts (e.g., Piran 2000), $\tilde L_\pm\sim
10^{50}-10^{52}$ ergs~s$^{-1}$, greatly exceed $L_\pm^{\rm max}$
for their estimated $\Gamma_\pm\sim 10^2$. For such a powerful
wind the pair density near the source is very high, and the
out-flowing pairs and photons are nearly in thermal equilibrium
almost up to the wind photosphere (e.g., Paczy\'nski 1990). The
outflow process of such a wind may be described fairly well by
relativistic hydrodynamics (Paczy\'nski 1986, 1990; Goodman 1986;
Grimsrud \& Wasserman 1998; Iwamoto \& Takahara 2002).
\par
In contrast if  $\tilde L_\pm \ll L_\pm^{\rm max}$ annihilation of
the outflowing pairs is negligible. It is now commonly accepted
that the magnetospheres of radio pulsars contain such a very
rarefied ultra-relativistic ($\Gamma_\pm \sim 10-10^2$) pair
plasma that is practically collisionless (see Melrose 1995 for a
review).

Recently we described a numerical code and the results of
calculations of spherically out-flowing, non-relativistic
($\Gamma_\pm\sim 1$) pair winds with total luminosity  in the
range $10^{34}-10^{42}$ ergs~s$^{-1}$, that is $L\sim
(10^{-2}-10^{6})L_\pm^{\rm max}$ (Aksenov, Milgrom, \& Usov 2004,
hereafter Paper~I). (A brief account of the emerging emission from
such a pair wind has been given by Aksenov Milgrom \& Usov 2003.)
While our numerical code can be more generally employed, the
results we presented in Paper~I were for a hot, bare, strange star
as the a wind injection source. Such stars are thought to be
powerful sources of pairs created by the Coulomb barrier at the
quark surface (Usov 1998, 2001a, see Fig.~1). For such
luminosities it is not justified to assume thermal equilibrium,
and so we have calculated the detailed microphysics including all
two-body processes ($ee\rightarrow ee$, $\gamma e\rightarrow
\gamma e$, $e^+e^-\leftrightarrow \gamma\gamma$) as well as their
radiative three-body variants ($ee\leftrightarrow ee\gamma$,
$\gamma e\leftrightarrow \gamma e\gamma$, $e^+e^-\leftrightarrow
\gamma\gamma\gamma$). The relativistic kinetic Boltzmann equations
for pairs and photons were solved numerically to describe the time
evolution and structure of pair winds and to find the emergent
emission (we only presented results for stationary situations).
Gravity of the star was neglected, and it was assumed that only
$e^\pm$ pairs are emitted from the stellar surface. Thermal
emission of photons from a bare quark surface, which has been
neglected in Paper I, is strongly suppressed relative to
black-body emission, but not completely, (e.g., Cheng \& Harko
2003; Jaikumar et al. 2004, see Fig.~1). Still, whatever little is
emitted may significantly affect the wind properties (see below).
\par
In the present paper we extend our previous studies to also
include the effects of gravity by solving the Boltzmann equations
for pairs and photons in a Schwarzschild space-time appropriate
for the strange star. We also include thermal emission of photons
from the stellar surface.
\par
 In \S~2 we formulate the equations that
describe the pair wind and the boundary conditions. In \S~3 we
describe the computational method used to solve these equations.
In \S~4 we present the results of our numerical simulations.
Finally, in \S~5, we discuss our results and some potential
astrophysical applications.

\section{Formulation of the problem}
We consider an $e^\pm$ pair wind that flows away from a hot, bare,
unmagnetized, non-rotating, strange star. Space-time outside the
star is described by  Schwarzschild's metric with the line element
\begin{equation}
  ds^2
 =-e^{2\phi}c^2
dt^2+e^{-2\phi}dr^2+r^2(d\theta^2+\sin^2\theta\, d\varphi^2)\,,
\label{ds1}
\end{equation}
where
\begin{equation}
  e^{\phi}=(1-r_g/r)^{1/2}
\end{equation}
is the lapse or redshift factor induced by gravity at a distance
$r$ from the stellar center,
\begin{equation}
r_g=2GM/c^2\simeq 2.95 \times 10^5 M/M_\odot\,\,{\rm cm}
\end{equation}
is the gravitational radius, with $M$ the gravitational mass of
the star.

Following Page \& Usov (2002) we consider, as a representative
case, a strange star with  $M=1.4 M_\odot$ which is constructed by
solving the Tolman-Oppenheimer- Volkoff equation of hydrostatic
equilibrium using an equation of state for SQM as described in
Alcock, Farhi, \& Olinto (1986a) and in Haensel, Zdunik, \&
Schaefer (1986) [with a bag constant $B = (140 {\rm MeV})^4$, QCD
coupling constant $\alpha_c \equiv g^2/4\pi = 0.3$, and the mass
of strange quarks $m_s = 150$ MeV]. The circumferential radius of
the star is $R =1.1\times 10^6$ cm.

We assume that the temperature is constant over the surface of the
star. The thermal emission in pairs and photons from the surface
of strange quark matter (SQM) depends on the surface temperature,
$T_{\rm S}$, alone (Usov 1998, 2001a; Cheng \& Harko 2003;
Jaikumar et al. 2004). The state of the plasma in the wind may be
described by the distribution functions $f_\pm ({\bf p}, r,t)$ and
$f_\gamma ({\bf p}, r,t)$ for positrons $(+)$, electrons $(-)$,
and photons, respectively, where ${\bf p}$ is the momentum of
particles. There is no emission of nuclei from the stellar
surface, so the distribution functions of positrons and electrons
are identical, $f_+=f_-$.
\par
Since we fix the mass and radius of the central star the only
remaining parameters that determine the wind are the surface
temperature, which determines the pair injection rate and
spectrum, and the photon surface radiation, which for default of
better knowledge we take to be thermal with a suppression factor,
$\eta$, relative to black body; $\eta$ is our second free
parameter. Because of the very steep dependence of the pair
injection rate on the surface temperature, we actually use the
former as a free parameter: in our calculations here $\tilde
L_\pm=10^{34},\,10^{35},\, ...\,,\, 10^{39}$ ergs~s$^{-1}$.

\subsection{General Relativistic Boltzmann Equations}
The Boltzmann equations in a Schwarzschild background for either
massless or massive particles can be obtained as a special case
from the results of Harleston \& Holcomb (1991) and Harleston \&
Vishniac (1992). These are given in a conservative form, which
makes them particularly amenable to numerical treatment:
\begin{eqnarray}
  \frac{e^{-\phi}}{c}\frac{\partial f_i}{\partial t}
 +\frac{1}{r^2}\frac{\partial}{\partial r}(r^2\mu e^{\phi}\beta_i
f_i)
 -\frac{e^{\phi}}{p^2}\frac{\partial}{\partial p}
  \left(
    p^3 \mu \frac{\phi'}{\beta_i} f_i
  \right) \nonumber \\
 -\frac{\partial}{\partial\mu}
  \left[
    (1-\mu^2)e^\phi
          \left(\frac{\phi'}{\beta_i}-\frac{\beta_i}{r}\right) f_i
  \right]
 =\sum_q(\bar\eta^q_i-\chi^q_i f_i).
\label{dfi}
\end{eqnarray}
Here, $i$ is the species type ($i=e$ for $e^\pm$ pairs and
$i=\gamma$ for photons), $\mu =\cos \theta$, $\theta$ is the angle
between the radius-vector ${\bf r}$ from the stellar center and
the particle momentum ${\bf p}$, $p=|{\bf p}|$,
$\beta_e=v_e/c=[1-(m_e c^2/\epsilon_e)^2]^{1/2}$, $\beta_\gamma
=1$, and $\epsilon_e=c[p^2+(m_ec)^2]^{1/2}$ is the energy of
electrons and positrons (for photons $\epsilon_\gamma=pc$). Also,
$\bar\eta_i^q$ is the emission coefficient for the production of a
particle of type $i$ via the physical process labelled by $q$, and
$\chi_i^q$ is the corresponding absorption coefficient. The
summation runs over physical processes that involve a particle of
type $i$.

For convenience of numerical simulations we use instead of $f_i$
the quantities
\begin{equation}
E_i(\epsilon,\mu,r,t)
   ={2\pi  \epsilon^3\beta_i f_i\over c^3},
\label{Ei}
\end{equation}
 standardly used in the ``conservative'' numerical method. This can
provide exact conservation of energy on a finite computational
grid (see below). $E_i$ is the energy density in the $\{{\bf
r},\,\mu,\,\epsilon\}$ phase space. Since all equations are the
same for all particle species from now on we omit the index $i$.

From equations (\ref{dfi}) and (\ref{Ei}) the Boltzmann equations
can be written in terms of $E$ as
\begin{eqnarray}
  \frac{e^{-\phi}}{c}\frac{\partial E}{\partial t}
 +\mu e^{-\phi}\frac{\partial}{r^2\partial r}
  (r^2\beta e^{2\phi} E)
 -\mu e^\phi\phi'
  \frac{\partial}{\partial\epsilon}
  (\epsilon\beta E) \nonumber \\
 - e^\phi\frac{\partial}{\partial\mu}
  \left[
  (1-\mu^2)
   \left(
     \frac{\phi'}{\beta}
     -\frac{\beta}{r}
   \right)
        E
  \right]
 =\sum_q(\eta^q-\chi^q E),\label{Boltzmann}
\end{eqnarray}
where
\begin{equation}
\eta^q={2\pi \epsilon^3\beta\bar\eta^q\over c^3}. \label{etai}
\end{equation}

\subsection{Boundary conditions}
The computational boundaries are
\begin{equation}
R< r < r_{\rm ext}, \,\,\,\,\,\,\,0 < t < t_{\rm st}\,,
\label{rint}
\end{equation}
where $r_{\rm ext}=1.66\times 10^8$ cm, and $t_{\rm st}$ is the
time when the wind approaches stationarity close enough, since
here we concentrate on steady-state winds.

Considering the interior of the quark star, it was shown that a
thin ($\sim 10^{-10}$~cm) layer of electrons with an extremely
strong electric field--the electrosphere--forms at the surface of
strange quark matter (SQM) (Alcock et al. 1986a; Kettner et al.
1995), which prevents the electrons from escaping to infinity.
 The electric field in the elecrosphere is a few ten times
higher than the critical field, $E_{\rm cr}\simeq 1.3\times
10^{16}$ V~cm$^{-1}$, at which the vacuum is unstable to creation
of $e^+e^-$ pairs \cite{S51}. Therefore, a hot, bare strange star
is potentially a powerful source of $e^+e^-$ pairs created in the
electrosphere and flowing away from the star (Usov 1998). We use
in our simulations a pair injection rate (Usov 2001a):
\begin{equation}
\dot N_\pm^{\rm in}=4\pi R^2F_\pm\,, \label{dotNep}
\end{equation}
where
\begin{eqnarray}
F_\pm =3\times 10^{42}
             \exp\,(- 0.593\zeta)
             \nonumber \\
             \times
             \left[\frac{\displaystyle\ln(1+2\zeta^{-1})}
                    {\displaystyle(1+0.074\zeta)^3}
+
                    \frac{\displaystyle\pi^5\zeta}
                    {\displaystyle2(13.9+\zeta)^4}
             \right]\,{\rm cm}^{-2}{\rm s}^{-1},
\label{Fpm}
\end{eqnarray}
and $\zeta=20(T_{\rm S}/10^9\,{\rm K})^{-1}$. The energy spectrum
of injected pairs is thermal with the surface temperature $T_{\rm
S}$, and their angular distribution is isotropic for $0\leq \mu
\leq 1$.

Thus, the pair injection luminosity, our first free parameter, is
given by
\begin{equation}
\tilde L_\pm = \dot N_\pm^{\rm in}[m_ec^2 +(3/2)k_{\rm B}T_{\rm
S}] \,, \label{tildeLpm}
\end{equation}
where $k_{\rm B}$ is the Boltzmann constant (Usov 2001a). For the
range  we consider, $\tilde L_\pm = 10^{34} - 10^{39}$
ergs~s$^{-1}$, the surface temperature is in a rather narrow
range, $T_{\rm S}\simeq (4-6)\times 10^8$~K (see Fig.~1).

Thermal emission of photons from the surface of a bare strange
star is strongly suppressed if the surface temperature is not very
high, $T_{\rm S}\ll 10^{11}$ K (Alcock et al. 1986a). The reason
is that the plasma frequency of quarks in SQM is very large,
$\hbar \omega_{\rm p}\simeq 20-25$ MeV, and only hard photons with
energies $\epsilon_\gamma>\hbar \omega_{\rm p}$ can propagate in
the SQM. The luminosity in such hard photons, which are in
thermodynamic equilibrium with quarks, decreases very fast for
$T_{\rm S} \ll \hbar \omega_{\rm p}/k_{\rm B}$ (Chmaj, Haensel, \&
Slomi\'nski 1991; Usov 2001a) and in our case, where $T_{\rm
S}\lesssim 10^9$ K, it is negligible (see Fig. 1). However,
low-energy ($\epsilon_\gamma <\hbar \omega_{\rm p}$) photons may
still leave the SQM if they are produced by  nonequilibrium
processes in the surface layer of thickness $\sim c/\omega_{\rm
p}\sim 10^{-12}$ cm (Chmaj et al. 1991). The emissivity of SQM in
nonequilibrium quark-quark bremsstrahlung radiation has been
estimated by Cheng \& Harko (2003) who find that it is suppressed
at least by a factor of $10^6$ in comparison with black-body
emission, $\tilde L_\gamma \lesssim 10^{-6}\, L_{\rm bb}$.  Usov
(2004) and Usov, Harko, \& Cheng (2005) have recently considered a
modified model of the electrosphere taking into account surface
effects for SQM . The modification of the electrosphere results in
an increase in the density of electrons by a factor of $\sim
20-30$ in comparison with the case when the surface effects are
ignored. The electron density increase can additionally suppress
the outgoing radiation from SQM in nonequilibrium quark-quark
bremsstrahlung photons. In our simulations we take
\begin{equation}
\tilde L_\gamma =\eta\, L_{\rm bb}= \eta\, 4\pi \sigma R^2T^4_{\rm
S}\,, \label{tLgamma}
\end{equation}
where $\eta$ is a dimensionless free parameter. We present results
with $\eta =0, 3\times 10^{-8}, 10^{- 6}$. In default of better
knowledge the photon spectrum is taken as black-body with the
surface temperature $T_{\rm S}$ corresponding to $\tilde L_\pm$.

An important parameter characterizing the affect of photon
emission from the surface on the out-flowing wind is the Eddington
luminosity $L_{\rm Edd}^\pm$ for the pair plasma, above which the
radiation pressure force dominates over gravity,
\begin{equation}
L_{\rm Edd}^\pm\simeq 10^{35}(M/1.4M_\odot )\,\, {\rm
ergs~s}^{-1}.
\end{equation}
Within the narrow range of surface temperatures studied here,
$T_{\rm S}\simeq (4-6)\times 10^8$~K, we get $\tilde L_\gamma$
that is $\sim (1-10)L^\pm_{\rm Edd}$ for $\eta =3\times 10^{-8}$,
and $\sim (10^2-10^3)L^\pm_{\rm Edd}$ for $\eta = 10^{-6}$ (see
Fig.~1).

The stellar surface is assumed to be a perfect mirror for both
$e^\pm$ pairs and photons. At the external boundary ($r=r_{\rm
ext}$) pairs and photons escape freely from the studied region,
i.e., the inward ($\mu < 0$) fluxes of both $e^\pm$ pairs and
photons vanish there.

\subsection{Physical processes in the pair plasma}
As the plasma moves outwards photons are produced by pair
annihilation ($e^+e^-\rightarrow \gamma\gamma$). Other two-body
processes that occur in the outflowing plasma, and are included in
the simulations, are M{\o}ller ($e^+e^+ \rightarrow e^+e^+ $ and
$e^-e^-\rightarrow e^-e^-$) and Bhaba ($e^+e^-\rightarrow e^+e^-$)
scattering, Compton scattering ($\gamma e\rightarrow \gamma e$),
and photon-photon pair production ($\gamma\gamma\rightarrow
e^+e^-$).

Two-body processes do not change the total number of particles in
the system, and thus cannot in themselves lead to thermal
equilibrium. So, we also include radiative processes
(bremsstrahlung, double Compton scattering, and three-photon
annihilation with their inverse processes), even though their
cross-sections are at least $\sim\alpha^{-1}\sim 10^2$ times
smaller than those of the two-body processes ($\alpha=e^2/\hbar
c=1/137$ is the fine structure constant).

\section{The computational method}
Our grid in the $\{{\bf r}, \mu,\epsilon\}$ phase-space is defined
as follows. The $r$ domain ($R<r<r_{\rm ext}$) is divided into
$j_{\rm max}$ spherical shells whose boundaries are designated
with half integer indices. The $j$ shell ($1\leq j\leq j_{\rm
max}$) is between $r_{j-1/2}$ and $r_{j+1/2}$, with $\Delta
r_j=r_{j+1/2}-r_{j-1/2}$ ($r_{1/2}=R$ and $r_{j_{\rm
max}+1/2}=r_{\rm ext}$).

The $\mu$-grid is made of $k_{\rm max}$ intervals
$\Delta\mu_k=\mu_{k+1/2}-\mu_{k-1/2}$: $1\leq k\leq k_{\rm max}$.

The energy grids for photons and electrons are both made of
$\omega_{\rm max}$ energy intervals $\Delta \epsilon_
{\omega}=\epsilon_{\omega +1/2}-\epsilon_{\omega-1/2}$:
$1\leq\omega \leq \omega_{\rm max}$, but the lowest energy for
photons is 0, while that for pairs is $m_ec^2$.

The quantities we compute are the energy densities averaged over
phase-space cells
\begin{equation}
    E_{\omega,k,j}(t)
   =\frac{1}{\Delta X}
    \int_{\Delta\epsilon_\omega,\Delta\mu_k,\Delta r_j}
    E\, d\epsilon\, d\mu\, r^2dr.
\end{equation}
where $\Delta X=\Delta \epsilon_\omega\Delta \mu_k \Delta
(r^3_j)/3$ and $\Delta (r^3_j)=r^3_{j+1/2}-r^3_{j-1/2}$.

Replacing the space, angle, and energy derivatives in the
Boltzmann equations (\ref{Boltzmann}) by finite differences (e.g.,
Mezzacappa \& Bruenn 1993) we have the following set of ordinary
differential equations (ODEs) for $E_{\omega,k,j}$ specified on
the computational grid:
\begin{eqnarray}
  \frac{e^{-\phi_j}}{c}
  \frac{d E_{\omega,k,j}}{d t}
 +e^{-\phi_j}\beta_{\omega}
  \frac{\Delta (r^2 \mu_k e^{2\phi}E_{\omega,k})_j}
  {\Delta (r_j^3)/3}
  \nonumber \\
 -\mu_k e^{\phi_j}\phi'_j
  \frac{\Delta(\epsilon\beta E_{k,j})_\omega}
  {\Delta\epsilon_\omega}
 -e^{\phi_j}
            \left(
            \frac{\displaystyle\phi'_j}{\displaystyle\beta_\omega}
           -\left<\frac{\displaystyle 1}{\displaystyle
r}\right>_j\beta_{\omega}
          \right)
  \nonumber \\
 \times\frac{\Delta\left[(1-\mu^2)
 E_{\omega,j}
        \right]_k}
  {\Delta\mu_k}
 =\sum_q[\eta_{\omega,k,j}^q-(\chi E)_{\omega,k,j}^q],
 \label{ODE}
\end{eqnarray}
where
\begin{equation}
\beta_{\omega}=\cases{1 &{for photons},\cr
[1-(m_ec^2/\epsilon_{\omega})^2]^{1/2}&{for electrons},\cr}
\end{equation}
\begin{equation}
  \epsilon_{\omega}
 =\frac{\epsilon_{\omega-1/2}+\epsilon_{\omega+1/2}}{2},
\end{equation}
\begin{equation}
  \mu_k=\frac{\mu_{k-1/2}+\mu_{k+1/2}}{2},
\end{equation}
\begin{equation}
  r_j=\frac{r_{1-1/2}+r_{j+1/2}}{2}.
\end{equation}
\begin{equation}
\phi_j=\phi(r_j)\,,\,\,\,\,\, \phi'_j=\phi'(r_j)\,
\end{equation}
\begin{equation}
  \left<\frac{\displaystyle 1}{\displaystyle r}\right>_j
 =\frac{\displaystyle (r_{j+1/2}^2-r_{j-1/2}^2)/2}{\displaystyle
  (r_{j+1/2}^3-r_{j-1/2}^3)/3},
\end{equation}
\begin{equation}
  E_{\omega,k}(r)
 =\frac{1}{\Delta\epsilon_\omega\Delta\mu_k}
  \int_{\Delta\epsilon_\omega\Delta\mu_k}
  E(\epsilon,\mu,r)
  d\epsilon d\mu,
\end{equation}
\begin{equation}
  E_{\omega,j}(\mu)
 =\frac{3}{\Delta\epsilon_\omega\Delta r_j^3}
  \int_{\Delta\epsilon_\omega\Delta r_j}
  E(\epsilon,\mu,r)
  drd\epsilon,
\end{equation}
\begin{equation}
  E_{k,j}(\epsilon)
 =\frac{3}{\Delta\mu_k\Delta r_j^3}
  \int_{\Delta\mu_k\Delta r_j^3}
  E(\epsilon,\mu,r)
  drd\mu\,,
\end{equation}
\begin{eqnarray}
   \Delta (r^2\mu_k e^{2\phi} E_{\omega,k})_j
  = r_{j+1/2}^2e^{2\phi_{j+1/2}}(\mu_k
E_{\omega,k})_{r=r_{j+1/2}}\nonumber\\
   -r_{j-1/2}^2e^{2\phi_{j-1/2}}(\mu_k E_{\omega,k})_{r=r_{j-
1/2}},
\end{eqnarray}
\begin{eqnarray}
    \Delta\left(\epsilon\beta E_{k,j}\right)_\omega
  =
\epsilon_{\omega+1/2}\beta_{\omega+1/2}(E_{k,j})_{\omega+1/2
}\nonumber\\
   -\epsilon_{\omega-1/2}\beta_{\omega-1/2}(E_{k,j})_{\omega-
1/2},
\end{eqnarray}
\begin{eqnarray}
    \Delta\left[(1-\mu^2) E_{\omega,j}\right]_k
  = (1-\mu_{k+1/2}^2)
(E_{\omega,j})_{\mu=\mu_{k+1/2}}\nonumber\\
   -(1-\mu_{k-1/2}^2) (E_{\omega,j})_{\mu=\mu_{k-1/2}}\,,
\end{eqnarray}
\begin{eqnarray}
  (E_{k,j})_{\omega+1/2}
 =\cases{
  E_{\omega+1,k,j} \cr
  +\frac{(E_{\omega+2,k,j}-E_{\omega+1,k,j})
   (\epsilon_{\omega+1/2}-\epsilon_{\omega+1})}
   {\epsilon_{\omega+2}-\epsilon_{\omega+1}},
  \cr \mu\geq0 \cr
  E_{\omega,k,j} \cr
              +\frac{(E_{\omega,k,j}-E_{\omega-
1,k,j})(\epsilon_{\omega+1/2}-\epsilon_\omega)}
              {\epsilon_\omega-\epsilon_{\omega-1}},
  \cr \mu<0,
  }
  \label{Eomega}
\end{eqnarray}
\begin{eqnarray}
  (E_{\omega,j})_{\mu=\mu_{k+1/2}}
 = E_{\omega,k,j} \nonumber \\
  +\frac{\Delta\mu_k(E_{\omega,k,j}-E_{\omega,k-1,j})}
   {\Delta\mu_{k-1}+\Delta\mu_k}\,,
   \label{Emu}
\end{eqnarray}

\begin{eqnarray}
   (\mu_k E_{\omega,k})_{r=r_{j+1/2}}
 = (1-\tilde\chi_{\omega,k,j+1/2})
   \nonumber \\
   \times
   \biggl( \frac{\displaystyle \mu_k+|\mu_k|}{\displaystyle 2}
           E_{\omega,k,j}
          +\frac{\displaystyle \mu_k-|\mu_k|}{\displaystyle 2}
           E_{\omega,k,j+1}
   \biggr)
   \nonumber \\
  +\tilde\chi_{\omega,k,j+1/2}\mu_k
   \frac{\displaystyle E_{\omega,k,j}+E_{\omega,k,j+1}}
   {\displaystyle 2}\,,
\end{eqnarray}
\begin{eqnarray}
\tilde\chi_{\omega,k,j+1/2}^{-1}=
  1
 +\frac{1}{\chi_{\omega,k,j}\Delta r_j}
 +\frac{1}{\chi_{\omega,k,j}\Delta r_{j+1}}.
\end{eqnarray}
The dimensionless coefficient $\tilde\chi$ is introduced to
describe correctly both optically thin and optically thick
computational cells by means of a compromise between the high
order method and the monotonic transport scheme without the
artificial viscosity (e.g., Richtmeyer \& Morton 1967; Mezzacappa
\& Bruenn 1993; Aksenov 1998). It is worth noting that if the
values of $(E_{k,j})_{\omega+1/2}$ and $(E_{\omega,j})_{\mu=
\mu_{k+1/2}}$ calculated by equations (\ref{Eomega}) and
(\ref{Emu}), respectively, are negative they are taken to be zero.

The right side terms of equation (\ref{ODE}) are
\begin{equation}
\eta_{\omega,k,j}^q
  =\frac{\displaystyle 1}
   {\displaystyle \Delta X}
   \int_{\Delta\epsilon_\omega, \Delta\mu_k, \Delta r_j}
   \eta^q d\epsilon\, d\mu\,
r^2 dr.
\end{equation}
\begin{equation}
(\chi E)_{\omega,k,j}^q
  =\frac{\displaystyle 1}
   {\displaystyle \Delta X}
   \int_{\Delta\epsilon_\omega,\Delta\mu_k,\Delta r_j}
   \chi E^q d\epsilon\, d\mu\,
r^2 dr.
\end{equation}
For the physical processes included in our simulations
expressions $\eta_{\omega,k,j}^q$ and $(\chi E)_{\omega,k,j}^q$
are calculated in Paper~I.

In paper~I, where gravity is neglected, particle number is
explicitly conserved when only two-body processes are taken into
account; particle production occurs only via three-body processes.
In the general relativistic scheme developed in this paper we were
not able to achieve exact number conservation. However, in all
runs for a stationary solution the change of the particle flux due
to computational effects is at most a few percents. Energy
conservation is still exactly satisfied.

There are several characteristic times in the system. Some are
related to particle reaction times, some to the time to reach
steady state. These times may greatly differ from each other,
especially at high luminosities ($\gtrsim 10^{38}$ ergs~s$^{-1}$)
when the pair wind is optically thick. The set of equations
(\ref{ODE}) is  then stiff: at least some eigenvalues of the
Jacobi matrix differ significantly from each other, and the real
parts of the eigenvalues are negative. In contrast to Mezzacappa
\& Bruenn (1993) we use Gear's method (Hall \& Watt 1976) to solve
the ODEs finite differences analog (\ref{ODE}) of the Boltzmann
equations. This high-order, implicit method was developed
especially to find a solution of stiff sets of ODE. To solve a
system of linear algebraic equations at any time step of Gear's
method we use the cyclic reduction method (Mezzacappa \& Bruenn
1993). The number of operations per time step is $\propto
(\omega_\mathrm{max}k_\mathrm{max})^3 j_\mathrm{max}$, which
increases rapidly with the increase of $\omega_{\rm max}$ and
$k_{\rm max}$. Therefore, the numbers of energy and angle
intervals have to be rather limited in our simulations.

Here we use an $(\epsilon,\mu ,r)$-grid with
$\omega_\mathrm{max}=13$, $k_\mathrm{max}=8$, and $j_{\rm
max}=100$. The discrete energies (in keV) of the $\epsilon$-grid
minus the rest mass of the particles are 0, 2, 27, 111, 255, 353,
436, 491, 511, 530, 585, 669, 766, and $\infty$. This gives a
denser grid at low energies and near the threshold of pair
production, $\epsilon =m_ec^2$. The $\mu$-grid is uniform, with
$\Delta\mu_k=2/k_\mathrm{max}=1/4$. The shell thicknesses are
geometrically spaced: $\Delta r_1=2\times 10^{-4}\mbox{ cm}$, and
$\Delta r_j=1.3\Delta r_{j-1}$.

To check the effects of grid coarseness we also performed test
computations with $k_{\rm max}=4$ and 6, and separately with
$\omega_{\rm max}=10$ and 11. We did not observe
changes
in the results.

\section{Numerical results}
In this section, we give the results for the properties of
spherically symmetric winds consisting of $e^\pm$ pairs and
photons. The pair injection luminosity $\tilde L_\pm$ and $\eta$
are the parameters in our simulations. The photon injection
luminosity $\tilde L_\gamma$ is determined by equation
(\ref{tLgamma}) where the surface temperature $T_{\rm S}$ relates
to the value of $\tilde L_\pm$ [see equations (\ref{dotNep}) -
(\ref{tildeLpm})]. We start from an empty wind, injecting both
pairs at a rate $10^{34}$ ergs~s$^{-1}$ and photons at a
corresponding rate. After a steady state is reached we start a new
run with this steady state as initial condition, increase the
energy injection rate in pairs by a factor of 10, wait for steady
state, and so on.
\par
 We next present the results for the structure of
the stationary winds and their emergent emission at the external
boundary.
\par
Figure~2 shows the mean optical depth $\tau_\gamma (r)$ for
photons, from $r$ to $r_{\rm ext}$. The wind as a whole is
optically thick $[\tau_\gamma (R)>1]$ for $\tilde L_\pm \gtrsim
10^{37}$ ergs~s$^{-1}$ irrespective of $\eta$. For $\tilde
L_\pm=10^{39}$ ergs~s$^{-1}$ the radius of the wind photosphere
$r_{\rm ph}$, determined by condition $\tau(r_{\rm ph})=1$, is
$\sim 4\times 10^6$ cm (see Fig.~2). The wind photosphere is
always deep inside our chosen external boundary ($r_{\rm ph}\ll
r_{\rm ext}$), justifying our neglect of the inward ($\mu < 0$)
fluxes at $r=r_{\rm ext}$.

Figures 3 and 4 show the pair number density ($n_e$) and the bulk
velocity of the pair plasma ouflow ($v_e^{\rm out}$),
respectively, as functions of the distance from the stellar
surface. For high luminosities ($\tilde L_\pm\gtrsim 10^{37}$
ergs~s$^{-1}$) both these  quantities hardly depend on $\eta$,
i.e., the pair wind structure is determined only by $\tilde
L_\pm$. For low luminosities ($\tilde L_\pm\lesssim 10^{35}$
ergs~s$^{-1}$), $n_e$ and $v_e^{\rm out}$ depend significantly on
$\eta$. In particular, for $\tilde L_\pm=10^{34}$ ergs~s$^{-1}$
near the stellar surface ($r-R\lesssim 10^5$ cm) the pair density
for $\eta=0$ is $\sim 6$ times higher than for $\eta = 10^{-6}$,
while the velocity of the pair plasma  is $\sim 6$ times smaller.
This is because  for a surface temperature $T_{\rm S}\simeq
4\times 10^8$~K, which corresponds to $\tilde L_\pm=10^{34}$
ergs~s$^{-1}$, the mean kinetic energy of pairs near the surface,
$(3/2)k_{\rm B}T_{\rm S}\simeq 8.3\times 10^{-8}$ ergs is about
half of the gravitational binding energy, so an atmosphere may
form. In addition, for low surface photon luminosities ($\eta
\lesssim {\rm a~few} \times 10^{-8}$) radiation pressure is too
weak to suppress the  plasma atmosphere that results from the
presence of gravity. So, an atmosphere does
 form,  and, in turn, is conducive to pair
annihilation. In contradistinction for $\eta =10^{-6}$
 radiation pressure does suppress the formation of an atmosphere,
  and the pair wind structure, i.e., the
runs of $n_e(r)$ and $v_e^{\rm out}(r)$, is very similar to the
case with no gravity and $\eta=0$.

Figure~5 shows the rates of outflow of $e^\pm$ pairs ($\dot
N_\pm$) and photons ($\dot N_\gamma$) through the surface as
functions of $r$. For high luminosities ($\tilde L_\pm\gtrsim
10^{37}$ ergs~s$^{-1}$) the rate of pair outflow $\dot N_\pm$
almost doesn't depend on $\eta$ and decreases significantly at the
distance
\begin{equation}
l_{\rm ann}\simeq 10\left({\tilde L_\pm\over
10^{39}\,{\rm ergs~s}^{-1}}\right)^{-1}\,{\rm cm}
\end{equation}
from the stellar surface because of pair annihilation (see the
estimate of $l_{\rm ann}$ in Paper~I). For low luminosities
($\tilde L_\pm\lesssim 10^{35}$ ergs~s$^{-1}$) the dependence of
$\dot N_\pm$ on $r$ is different for different values of $\eta$.
For $\eta =0$  the value of $\dot N_\pm$ decreases outwards at
least by a factor of 2 even when $\tilde L_\pm$ is as small as
$10^{34}$ ergs~s$^{-1}$, while for $\eta= 10^{-6}$ the decrease of
$\dot N_\pm$ due to pair annihilation is very small for low
luminosities, again reflecting the degree to which a pair
atmosphere is formed.

The photon outflow rate is seen to increase with increasing $r$
not only because of pair annihilation, but also because of
accumulative production of photons by radiative three-body
processes. These processes are important for high luminosities
($\tilde L_\pm\gtrsim 10^{38}$ ergs~s$^{-1}$) and are responsible
for the increase of $\dot N_\gamma$ at $r-R\sim 10^3-10^5$ cm for
such high luminosities (see Fig.~5).

The rates of energy outflow in $e^\pm$ pairs ($\dot E_\pm$) and
photons ($\dot E_\gamma$) vary with radius more or less similarly
to the particle outflow rates, except that the total energy rate
is explicitly conserved in all processes (see Fig.~6). The total
energy rate decreases with increase of $r$ because of the gravity
effects,
\begin{equation}
\dot E(r) =\dot E_\pm (r) +\dot E_\gamma (r)=
e^{2(\tilde\phi -\phi)}\tilde L
={{1 -r_g/R}\over {1 - r_g/r}}\tilde L\,,
\end{equation}
and varies from $\tilde L=\tilde L_\pm +\tilde L_\gamma$
at $r=R$ to the total emerging luminosity
\begin{equation}
L=L_\pm+L_\gamma =\dot E(r_{\rm ext})
={{1 -r_g/R}\over {1 - r_g/r_{\rm ext}}}\tilde L
\end{equation}
at $r=r_{\rm ext}$. This differs qualitatively
from Paper~I
where the gravity effects are ignored, and $\dot E(r)$
is constant.

The number rates of emerging pairs $(\dot N_\pm)$ as functions of
$\tilde L_\pm$ for different values of $\eta$ are shown in
Figure~7. For $\tilde L_\pm=10^{34}$ ergs~s$^{-1}$ the value of
$\dot N_\pm$ for $\eta=0$ is $\sim 3$ times smaller than the same
for $\eta =10^{-6}$. The results of our new simulations for $\eta
=10^{-6}$ and our old simulations with $\eta=0$ and no gravity
practically coincide. This is consistent with our results on the
wind structure.  For $\tilde L_\pm \gtrsim 10^{37}$ ergs~s$^{-1}$
$\dot N_\pm$ hardly depends on $\eta$, and is $\sim 1.5-2$ times
smaller than the result of Paper~I (see Fig.~7). This is due to
partial suppression of pair creation as the photon energies are
reduced by gravitational redshift.

Figure~8 shows the emerging total luminosities in $e^\pm$ pairs
($L_\pm$, including the rest mass) and photons ($L_\gamma$)--given
as fractions of the total emerging luminosity
$L=L_\pm+L_\gamma$--as functions of the injection luminosity, and
for different values of $\eta$. For the luminosity range we
consider photons dominate in the emerging emission ($L_\gamma >
L_\pm$), especially at high luminosities where $L_\gamma\gg
L_\pm$.

Figure~9 presents the energy spectra of the emerging photons for
different values of $\tilde L_\pm$ and $\eta$. For low
luminosities ($\tilde L_\pm \lesssim 10^{36}$ ergs~s$^{-1}$) and
$\eta=0$, photons, which form by pair annihilation, escape more or
less freely from the star's vicinity, and the photon spectra
represent a very wide annihilation line that is redshifted in the
gravitational field. The mean energy of the emerging photons
varies from $\sim 430$ keV for $\tilde L_\pm =10^{34}$
ergs~s$^{-1}$ to 400 keV for $\tilde L_\pm=3\times 10^{36}$
ergs~s$^{-1}$ (see Fig.~10). If $\eta$ is near its maximum value
($\sim 10^{-6}$) there is no annihilation line in the spectra of
emerging photons. Annihilation photons only modify these spectra
by producing high-energy tails (see Fig.~9c). For high
luminosities ($\tilde L_\pm\gtrsim 10^{38}$ ergs~s$^{-1}$) the
energy spectra of emerging photons practically don't depend on
$\eta$ and coincide with the results of Paper~I.

\section{Discussion}

We have identified certain characteristics of pair winds
outflowing from hot, bare, strange stars and their emerging
emission. For the energy injection rate in pairs we consider
($\tilde L_\pm=10^{34}-10^{39}$ ergs~s$^{-1}$) photons dominate in
the emerging emission (see Fig.~8.). The spectrum of the emerging
photons, we find, is rather hard (see Figs. 9 and 10) and differs
qualitatively from the spectrum of the thermal emission from a
neutron star with the same luminosity (e.g., Romani 1987; Shibanov
et al. 1992; Rajagopal \& Romani 1996; Page et al. 2004). In
particular, for a bare strange star the mean energy of the
emerging photons (see Fig.~10) is at least an order of magnitude
larger than the same for a neutron star. This opens observational
possibilities to distinguish strange stars from neutron stars.
Hard spectra of bare strange stars are amenable to detection and
study by sensitive, hard X-ray and soft $\gamma$-ray instruments,
such as {\it INTEGRAL} (e.g., Winkler et al. 2003).

In this study we take into account both the thermal emission of
photons from the strange star surface and gravity that have been
neglected in Paper~I. We have shown that for low values of
$\eta\lesssim {\rm a~few} \times 10^{-8}$ and $\tilde
L_\pm\lesssim 10^{35}$ ergs~s$^{-1}$ pairs emitted by the stellar
surface are mainly captured by the gravitational field, and a pair
 atmosphere forms. The probability of pair annihilation
increases because of the increase of the pair number density in
the atmosphere, and this results in decrease of the fraction of
pairs in the emerging emission in comparison with the case of
Paper~I (see Figs. 7 and 8). However, if the surface emission in
photons is as high as its upper limit ($\eta \simeq 10^{-6}$)
radiation pressure forces dominate over gravity, and the pair wind
structure practically coincides with the results for no gravity,
irrespective of $\tilde L_\pm$.

The dominant physical processes in pair winds change significantly
depending on $\tilde L_\pm$. For $\tilde L_\pm < L_*\simeq
10^{37}$ ergs~s$^{-1}$ the optical depth for photons is smaller
than unity, $\tau_\gamma <1$ (see Fig.~2). In this case some part
of $e^\pm$ pairs ejected from the stellar surface annihilates into
photons in the process of their outflow, while photo-creation of
pairs is rather rare. Above $\tilde L_\pm \simeq L_*$  pair
creation by photons becomes important, and for $\tilde L_\pm \gg
L_*$ the rates of pair creation and pair annihilation are more or
less comparable at $l_{\rm ann}\lesssim r-R \ll r_{\rm ph}$. At
such distances from the surface the number density of pairs is
nearly constant (see Fig.~3 and Fig.~8 in Paper~I). In addition,
for high luminosities ($\tilde L_\pm \gtrsim 10^{38}$
ergs~s$^{-1}$) radiative three-body processes are important, and
the total rate of the particle outflow increases with radius (see
Fig.~5). These processes favour thermalization of pairs and
photons in the outflow.

For low luminosities ($\tilde L_\pm < L_*$) the spectrum of
emerging photons significantly depends on the photo-emission from
the surface, especially at low energies, $\epsilon_ \gamma\ll
m_ec^2$ (see Fig.~9). If surface photo-emission is negligible
($\eta =0$), this spectrum resembles a very wide annihilation
line. The fractional emerging luminosity in the annihilation line
decreases with the increase of $\eta$. For $\eta =10^{-6}$ the
annihilation line is completely washed out, and a high-energy part
of this line is observed as a high-energy tail (see Fig.~9c). For
high luminosities ($\tilde L_\pm \gtrsim 10^{38}$ ergs~s$^{-1}$)
the spectrum of emerging photons is practically independent of the
photon emission from the stellar surface.

Soft $\gamma$-ray repeaters (SGRs), which are the sources of short
bursts of hard X-rays with super-Eddington luminosities (up to
$\sim 10^{42}-10^{47}$ ergs~s$^{-1}$), are potential candidates
for strange stars (e.g., Alcock et al. 1986b; Cheng \& Dai 1998;
Usov 2001c; Ouyed et al. 2005). The bursting activity of SGRs may
be explained by fast heating of the stellar surface up to a
temperature of $\sim (1-3)\times 10^9$~K and its subsequent
thermal emission (Usov 2001b,c). The heating mechanism may be
either impacts of comets onto bare strange stars (Zhang et al.
2000; Usov 2001b) or fast decay of superstrong ($\sim
10^{14}-10^{15}$~G) magnetic fields (Thompson \& Duncan 1995; Heyl
\& Kulkarni 1998). Since for high luminosities ($\gtrsim 10^{38}$
ergs~s$^{-1}$) the spectrum of emerging photons practically
doesn't depend on the surface photo-emission, which is poorly
known, it is possible to calculate more securely many properties
(light curves, energy spectra, etc.) of powerful ($L\gg 10^{38}$
ergs~s$^{-1}$) X-ray bursts expected from the stellar surface
heating neglecting surface photo-emission. We plan to apply the
tools developed here to study these problems.
\begin{acknowledgements}
We thank members of the Israeli Center for High Energy Astrophysics,
D. Eichler, A. Levinson, T. Piran, and E. Waxman, for helpful
discussions. The research was supported by the Israel Science
Foundation of the
Israel Academy of Sciences and Humanities.
\end{acknowledgements}


\clearpage

\begin{figure}
\plotone{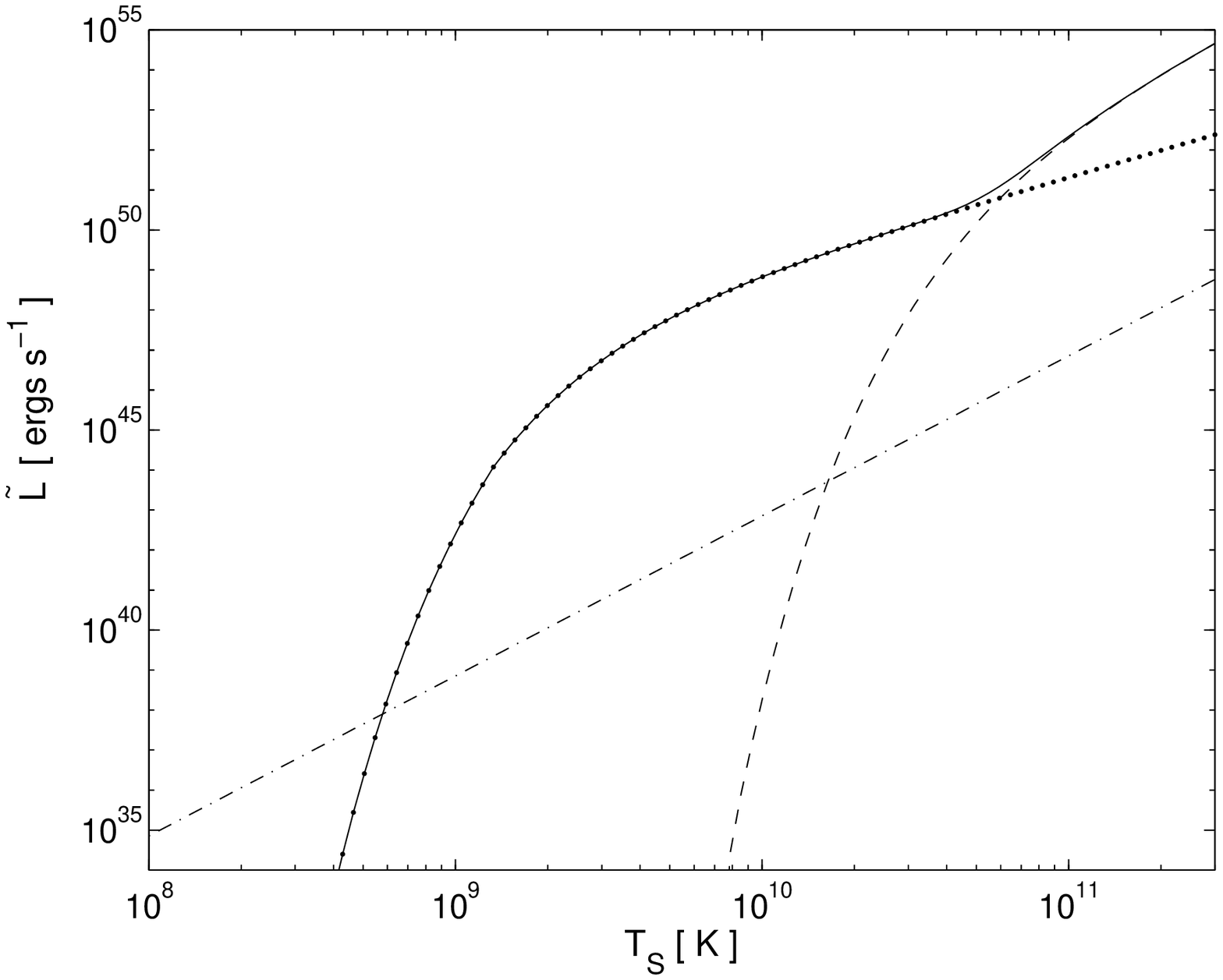} \caption{Injection luminosities of a hot, bare,
strange star in $e^+e^-$ pairs (dotted line), in thermal
equilibrium photons (dashed line), and the total (solid line) as
functions of the surface temperature $T_{_{\rm S}}$. The
theoretical upper limit on the luminosity in non-equilibrium
photons, $10^{-6}L_{\rm BB}$ (Cheng \& Harko 2003), is shown by
the dot-dashed line, $L_{\rm BB}$ being the blackbody luminosity.
} \label{Ltotal-Ts}
\end{figure}

\clearpage

\begin{figure}
\plotone{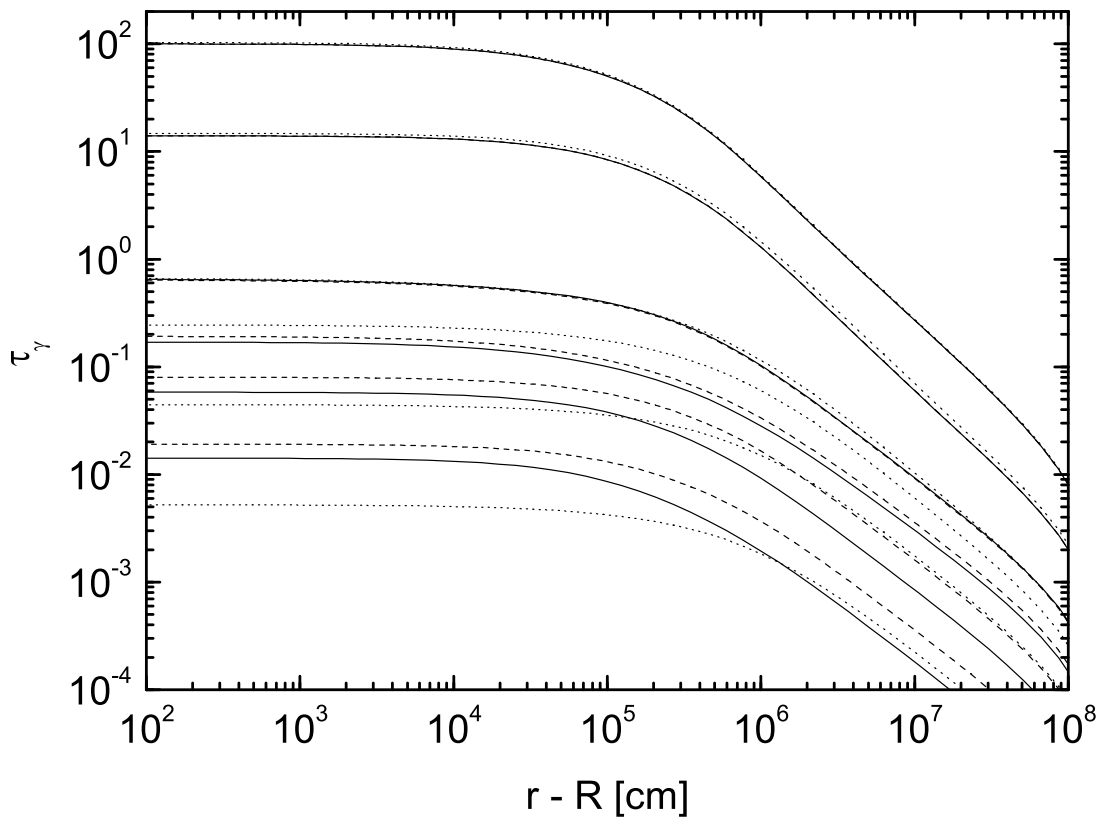} \caption{Mean optical depth for photons, from $r$
to infinity, as a function of the distance from the stellar
surface for $\eta =0$ (solid lines), $\eta =3\times 10^{-8}$
(dashed lines), and $\eta =10^{-6}$ (dotted lines), and for
different values of the injected pair luminosity, $\tilde L_\pm$,
which increases in steps of factor ten from $10^{34}$
ergs~s$^{-1}$ (lowest triplet of lines) to $10^{39}$ ergs~s$^{-1}$
(uppermost triplet of practically coinciding lines).}
\label{tau.Fig}
\end{figure}

\clearpage

\begin{figure}
\plotone{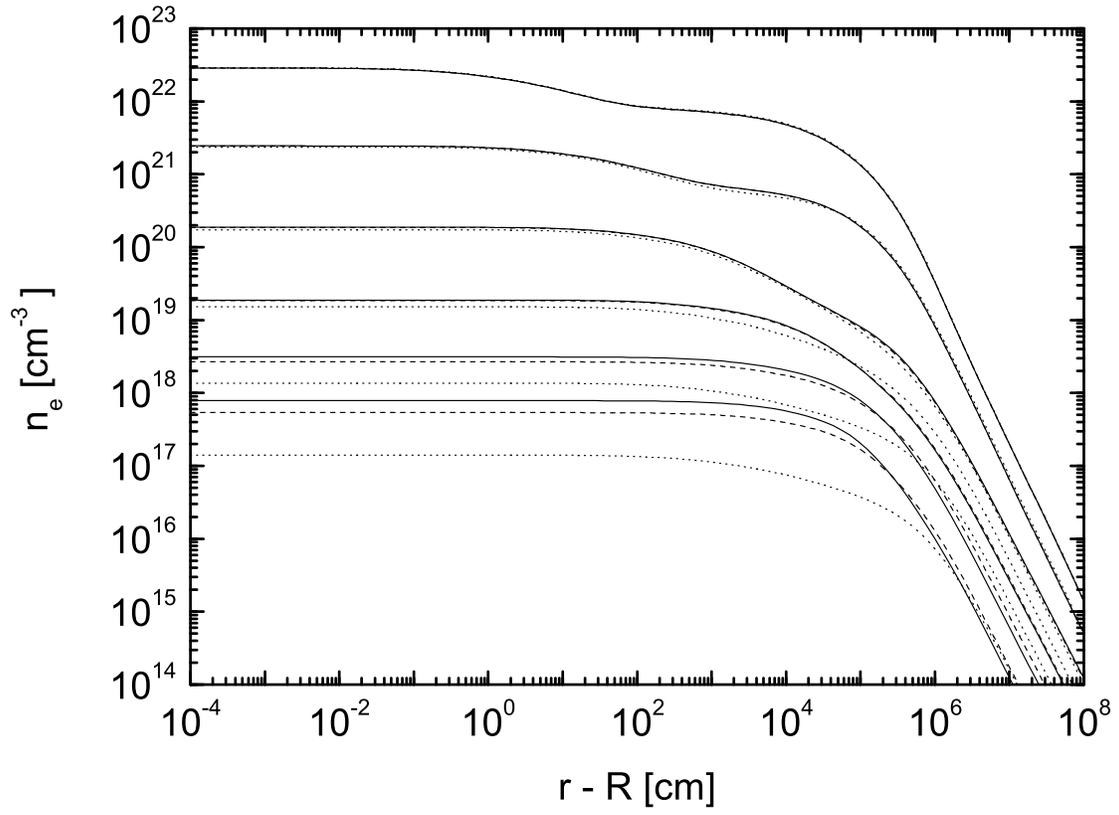} \caption{Pair number density as a function of the
distance from the stellar surface, line designation as in Fig. 2.}
\end{figure}

\clearpage

\begin{figure}
\begin{center}\resizebox{!}{0.8\textheight}{{\includegraphics{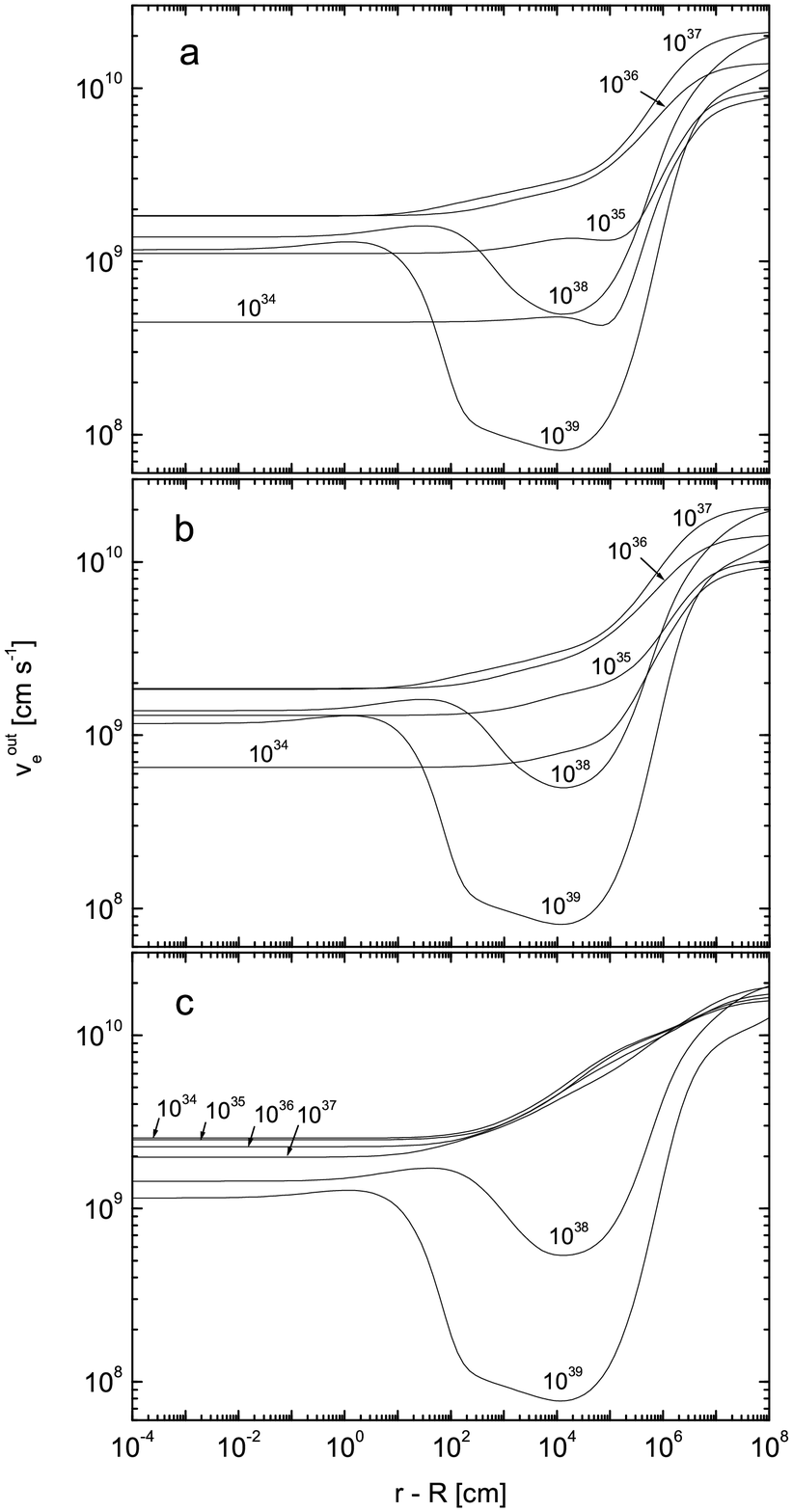}}}
\end{center} 
\caption{Bulk velocity of the pair plasma as a function of
the distance from the stellar surface for (a)
$\eta =0$, (b) $\eta= 3\times 10^{-8}$, and (c) $\eta = 10^{-6}$,
shown for different values of
$\tilde L_\pm$, as marked on the lines.} \label{v-gamma}
\end{figure}

\clearpage

\begin{figure}
\begin{center}\resizebox{!}{0.8\textheight}{{\includegraphics{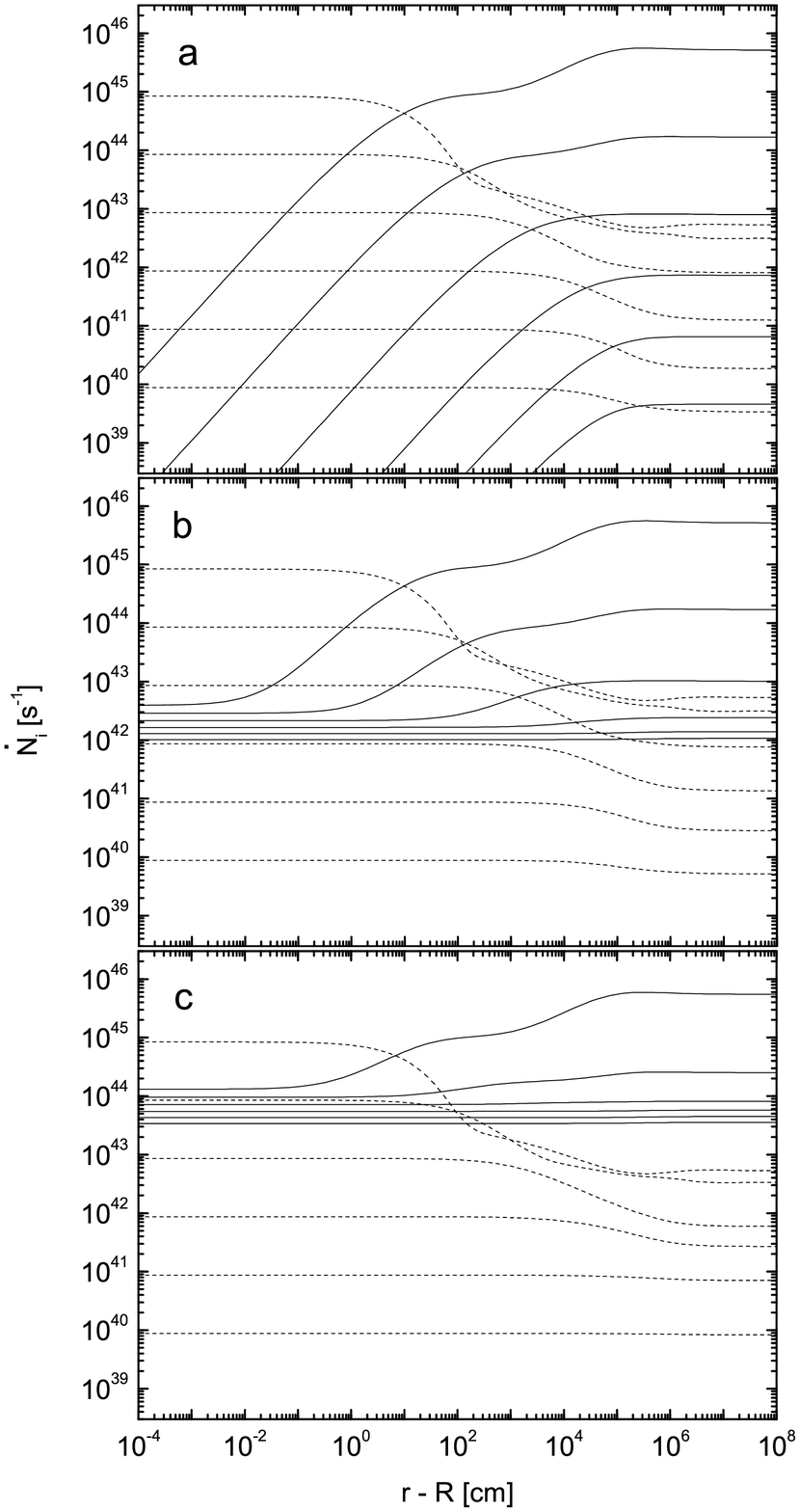}}}
\end{center}
\caption{Particle number outflow rates in photons (solid lines)
and $e^\pm$ pairs (dashed lines)
as functions of the distance from the stellar surface for (a)
$\eta =0$, (b) $\eta= 3\times 10^{-8}$, and (c) $\eta = 10^{-6}$,
shown for different values
of $\tilde L_\pm$,
which increases in steps of factor ten from
$10^{34}$ ergs~s$^{-1}$ (lowest
lines for each species)
to $10^{39}$ ergs~s$^{-1}$ (uppermost lines). }
\end{figure}

\clearpage

\begin{figure}
\begin{center}\resizebox{!}{0.8\textheight}{{\includegraphics{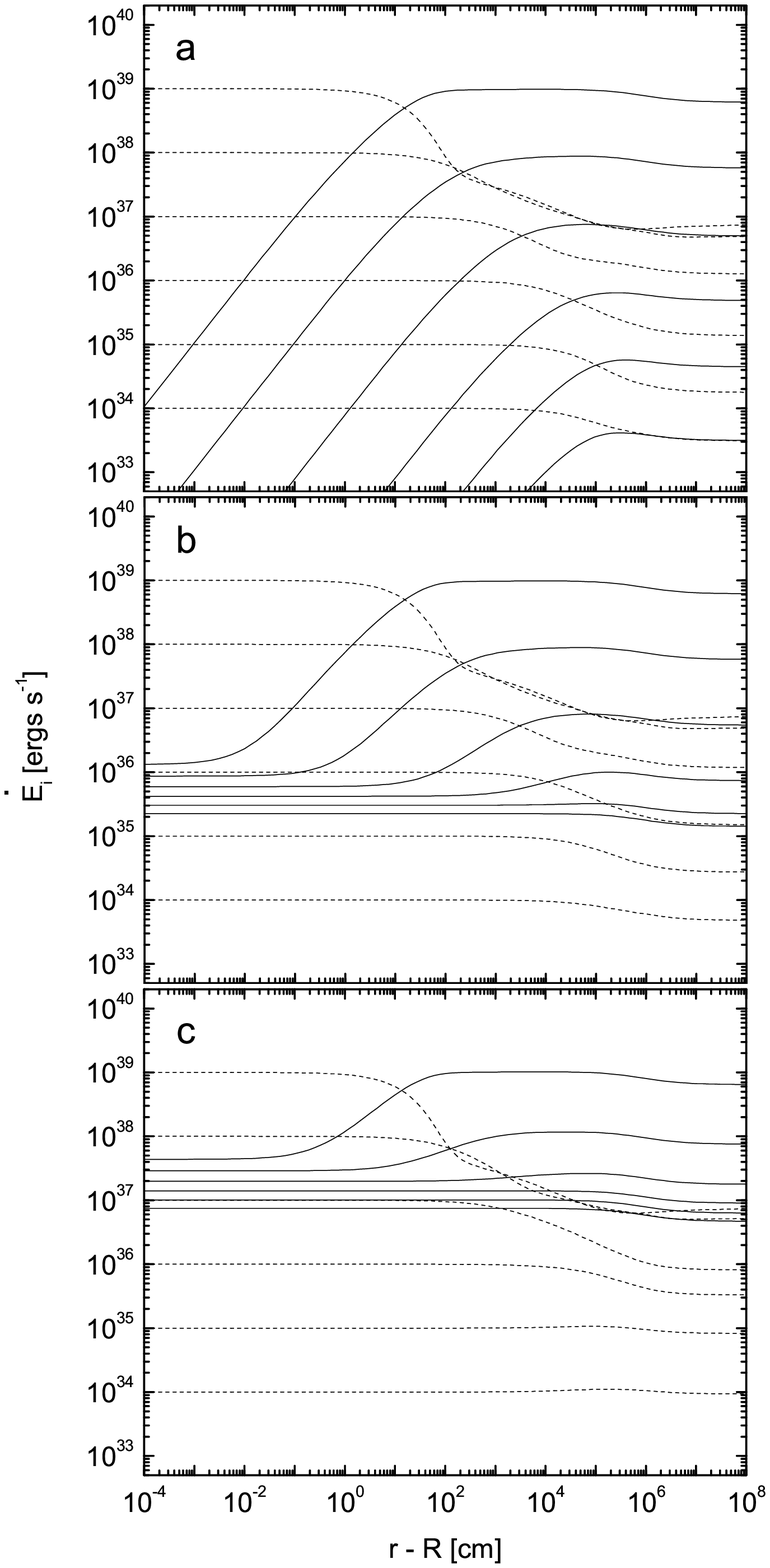}}}
\end{center} 
\caption{Rates of energy outflow, as in Fig. 5.}
\end{figure}

\clearpage

\begin{figure}
\plotone{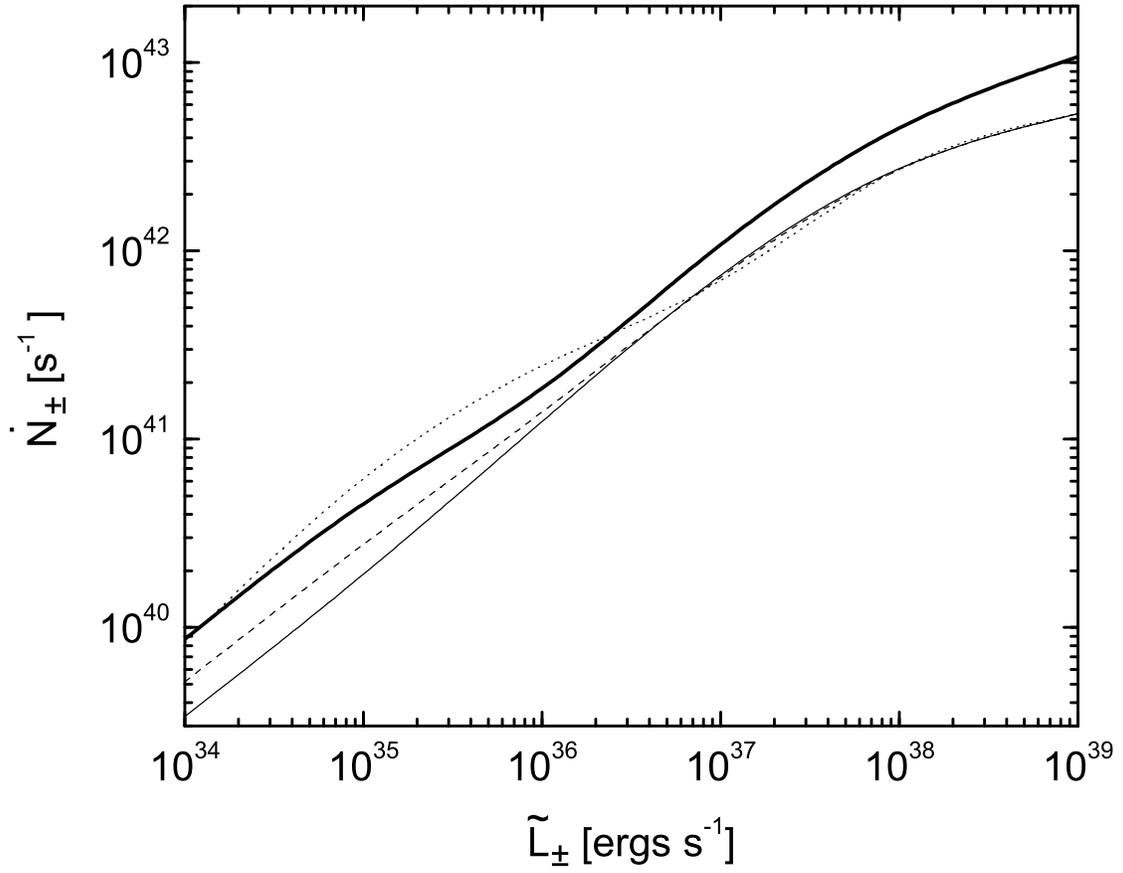} \caption{Number rates of emerging pairs as
functions of the injected pair luminosity for
$\eta =0$ (thin solid line), $\eta =3\times 10^{-8}$ (dashed
line), $\eta =10^{-6}$ (dotted line). The result of Paper~I
where gravity has been neglected and $\eta=0$ is shown by
the thick solid line.}
\end{figure}

\clearpage

\begin{figure}
\plotone{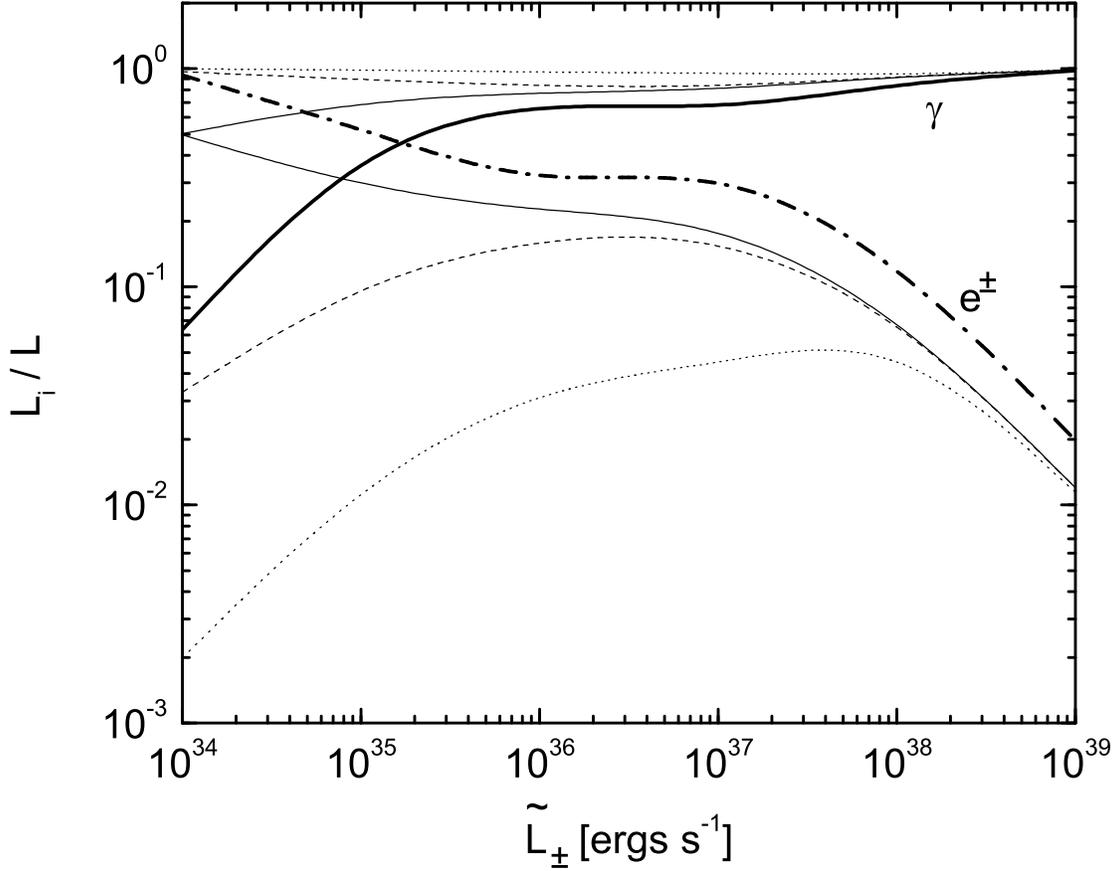} \caption{Fractional emerging luminosities in
photons (upper triplet of thin lines) and pairs (lower triplet of
thin lines) as functions of the injected pair luminosity for $\eta
=0$ (thin solid lines), $\eta =3\times 10^{-8}$ (dashed lines),
and $\eta =10^{-6}$ (dotted lines). The results for the case where
gravity is neglected and $\eta =0$ are shown for comparison by
thick solid line and thick dot-dashed line for emerging photons
and emerging pairs, respectively.}
\end{figure}

\clearpage

\begin{figure}
\begin{center}\resizebox{!}{0.8\textheight}{{\includegraphics{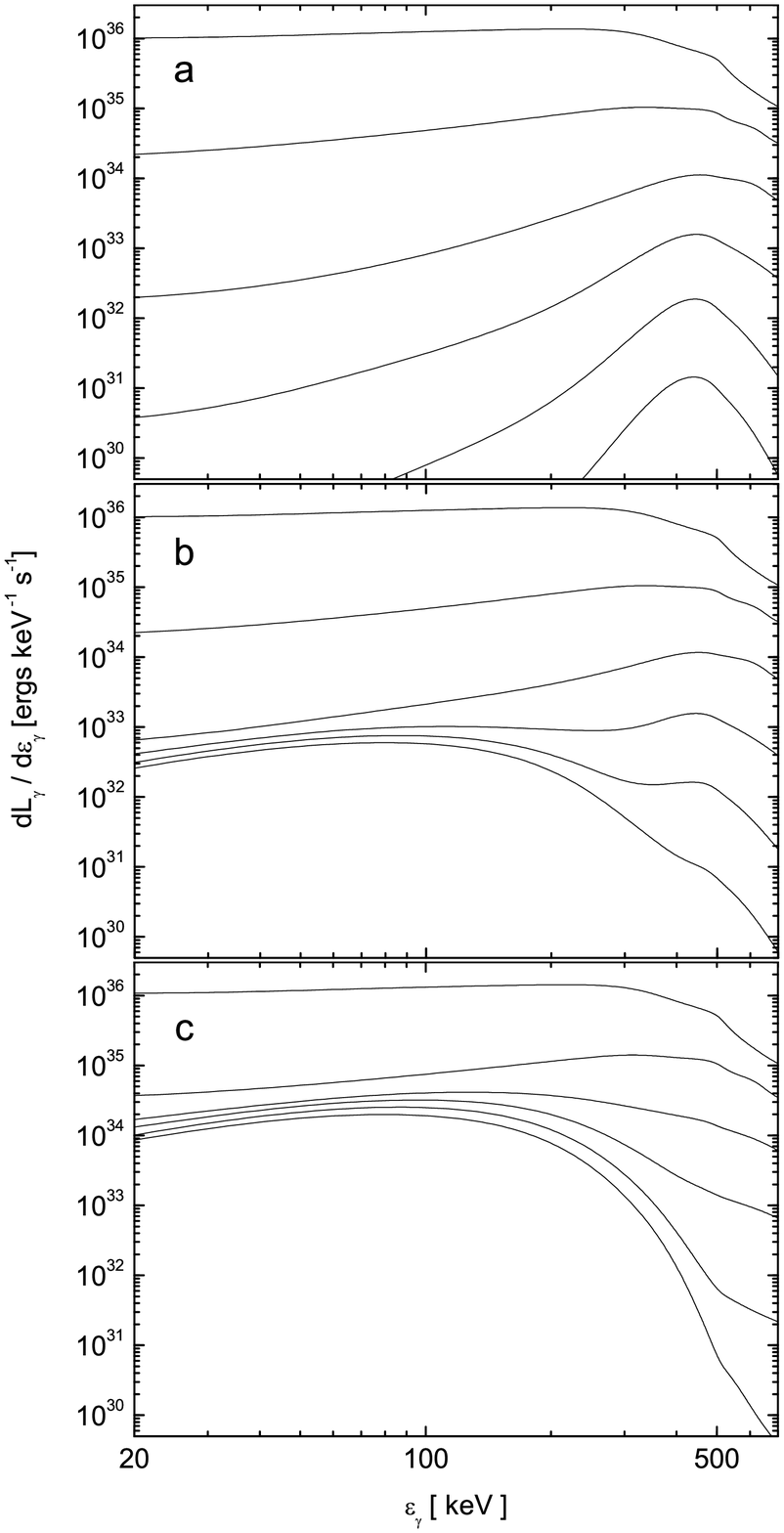}}}
\end{center} 
\caption{Energy spectrum of emerging photons for (a)
$\eta =0$, (b) $\eta= 3\times 10^{-8}$, and (c) $\eta = 10^{-6}$,
shown for different values
of $\tilde L_\pm$,
which increases in steps of factor ten from
$10^{34}$ ergs~s$^{-1}$ (lowest
lines) to $10^{39}$
ergs~s$^{-1}$ (uppermost lines).
}
\end{figure}

\clearpage

\begin{figure}
\plotone{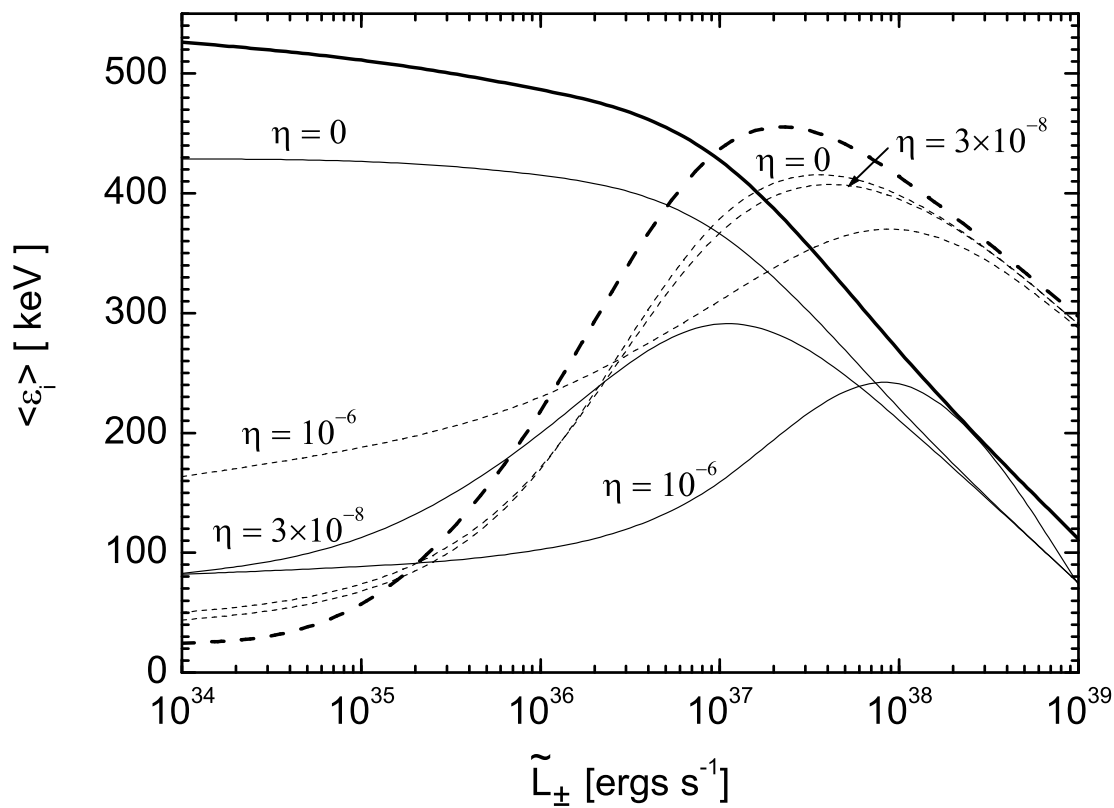} \caption{Mean energy of the emerging photons
(thin solid lines) and electrons (thin dashed lines)
as functions of the injected pair luminosity for
different values of $\eta$, as marked on the lines.
The results of Paper~I where $\eta=0$ and no gravity
are shown by the thick solid and thick dashed lines
for photons and electrons, respectively.}
\end{figure}

\end{document}